\documentclass[iop]{emulateapj}

\newcommand\cs{c_s}

\newcommand\Msun{\; {\rm M}_{\odot}}

\newcommand\kms{\; {\rm km}\;{\rm s}^{-1}}
\newcommand\pc{\;{\rm pc}}

\newcommand\kpc{\;{\rm kpc}}

\newcommand\freq{\kms\kpc^{-1}}

\newcommand\Myr{\;{\rm Myr}}
\newcommand\Gyr{\;{\rm Gyr}}

\newcommand\Surf{\Msun\;{\rm pc^{-2}}}

\newcommand\Omb{\Omega_{b}}

\newcommand\Qb{Q_b}

\newcommand\simgt{\lower.5ex\hbox{$\; \buildrel > \over \sim \;$}}
\newcommand\simlt{\lower.5ex\hbox{$\; \buildrel < \over \sim \;$}}

\def\spose#1{\hbox to 0pt{#1\hss}}
\def\dt{\spose{\raise 1.0ex\hbox{\hskip2pt$\mathchar"201$}}}

\slugcomment{Accepted to appear in the Astrophysical Journal}

\shorttitle{Seeding massive BHs in galaxies}
\shortauthors{Li, Sellwood \& Shen}

\begin{document}

\title{Rapid formation of black holes in galaxies: a self-limiting growth mechanism}

\author{Zhi Li\altaffilmark{1,2,5}, J. A. Sellwood\altaffilmark{3,4,6} and Juntai Shen\altaffilmark{1,2,7}}

\affil{$^1$Key Laboratory for Research in Galaxies and Cosmology, Shanghai Astronomical Observatory, Chinese Academy of Sciences, \\ 80 Nandan Road, Shanghai 200030, China \\
$^2$University of China Academy of Sciences, 19A Yuquan Road, Beijing 100049, China\\
$^3$Department of Physics and Astronomy, Rutgers University, 136 Frelinghuysen Road, Piscataway, NJ 08854\\
$^4$Current address: Steward Observatory, University of Arizona, 933 N Cherry Ave, Tucson AZ 85721\\
Emails:$^5$lizh@shao.ac.cn $^6$sellwood@physics.rutgers.edu $^7$jshen@shao.ac.cn}


\begin{abstract}
We present high-quality fluid dynamical simulations of isothermal gas
flows in a rotating barred potential.  We show that a large quantity
of gas is driven right into the nucleus of a model galaxy when the
potential lacks a central mass concentration, but the inflow stalls at
a nuclear ring in comparison simulations that include a central
massive object.  The radius of the nuclear gas ring increases linearly
with the mass of the central object.  We argue that bars drive gas
right into the nucleus in the early stages of disk galaxy formation,
where a nuclear star cluster and perhaps a massive black hole could be
created.  The process is self-limiting, however, because inflow stalls
at a nuclear ring once the mass of gas and stars in the nucleus
exceeds $\sim1$\% of the disk mass, which shuts off rapid growth of
the black hole.  We briefly discuss the relevance of these results to
the seeding of massive black holes in galaxies, the merger model for
quasar evolution, and the existence of massive black holes in disk
galaxies that lack a significant classical bulge.
\end{abstract}

\keywords{%
  galaxies: ISM ---
  galaxies: kinematics and dynamics ---
  galaxies: structures ---
  galaxies: hydrodynamics ---
  galaxies: formation ---
  quasars: supermassive black holes 
}

\section{Introduction}
Most galaxies host massive black holes in their centers
\citep{kor_ric_95}, although many are dormant.  Black holes are
believed to power quasars, which are thought to have short lifetimes
in the bright phase \citep{martin_04}.  It is well established that
the co-moving space density of quasars peaks at intermediate redshift
\citep[e.g.][]{boyle_etal_87,hopkin_etal_07a,singal_etal_16}.  The
rise of this activity from the time of the early universe to $z\sim3$
coincides with galaxy assembly that is driven by gravitational
instability, when gas is abundant \citep[see][for a recent
  review]{som_dav_15}.  The usual idea is that most quasar activity is
caused by galaxy mergers, which drive gas into the nucleus
\citep{too_too_72, hopkin_etal_07b, treist_etal_12, bonoli_etal_14}.
The decline in the quasar luminosity function since $z\sim2$ is
attributed to a variety of factors \citep{mer_hei_08}: the decrease in
the galaxy merger rate as the universe expands, a decrease in the gas
content of galaxies \citep[e.g.][]{mor_bab_15}, and feedback
\citep{wyith_etal_03,dimatt_etal_05}, although exactly how feedback
extinguishes activity is difficult to model
\citep[e.g.][]{hopkin_etal_05, dubois_etal_12}.

However, the merger model for quasar activity implicitly pre-supposes
the pre-existence of moderate mass black holes in the merging
galaxies.  Furthermore, stars formed prior to a galaxy merger are
expected to accumulate into a classical bulge \citep[but see][for a
  dissenting view]{kes_nus_12}, which is absent in some significant
fraction of galaxies \citep{kormen_etal_10}.  None of the galaxies
lacking classical bulges listed by \citet{kormen_etal_10} have
measured black hole masses but some have an AGN that is a clear
indicator of a moderately massive black hole, hereafter
MBH.\footnote{We use MBH to indicate BHs with masses over a broad
  range up to $\sim 10^7$M$_\odot$.}.  Two specific examples are (a)
NGC 5746, which has no classical bulge \citep{bar_kor_12} but has an
X-ray bright Seyfert nucleus \citep{gonzal_etal_09}, and (b) the dwarf
galaxy RGG 118 that has a pseudobulge and an AGN, for which
\citet{Baldass_17} estimate a BH mass of $\sim 5\times
10^4~$M$_\odot$.  A path must therefore exist to form a MBH in a
galaxy that does not involve mergers.

The formation of seed BHs has received a lot of attention
\citep[see][for recent reviews]{lat_fer_16,smith_etal_17}, but a
convincing model remains elusive.  Some have argued
\citep[e.g.][]{mad_ree_01} that the seeds are stellar mass BHs in the
early universe.  A second idea is runaway collapse of star clusters
\citep[e.g.][]{sha_teu_85, ebisuz_etal_01}.  The third suggestion,
which continues to be intensively studied, is the direct collapse of a
gas cloud to form a BH with a seed mass of $\sim 10^4$--$10^7\Msun$
\citep[e.g.][]{hae_ree_93,luo_etal_16}.

\citet[][hereafter SM99]{sel_mor_99} presented a little known
alternative scenario.  They suggested that MBHs could be created at
the centers of forming galaxies, where gas is concentrated into a
small volume through the action of a rotating bar.  We here review the
ingredients of their model, which attempts to account both for the
creation of MBHs and their subsequent quiescence.  Similar ideas were
also advanced independently in a much later paper by
\citet{fanali_etal_15}.

The model proposed by SM99 relates to the period of disk galaxy
assembly.  It starts from the well-known finding \citep{ost_pee_73,
  toomre_81, ber_sel_16} that the formation of a largely rotationally
supported galactic disk would naturally lead to the early formation of
a bar.  SM99 argued that the initial absence of a central mass
concentration would allow gas to be driven inward as far as the bar
torque could achieve.  They then speculated that a fraction of the gas
that accumulates in the center may collapse to grow and power a MBH,
while the remainder contributes to a central mass concentration of
stars and gas.\footnote{Hereafter, we refer to the combined mass of
  gas, stars, and the MBH that reside in the center as a ``massive
  central object.''}  They concluded that inflow driven by the early
bar would have built up a sufficiently massive central object that the
later gas inflow would stall at distances of a few hundred parsec from
the nucleus.  In their picture, the early formation of a bar grows the
MBH, but the associated nuclear activity subsides soon thereafter
because the build up of the central mass causes the inflow to stall at
a nuclear ring, thereby halting the rapid growth of the MBH.

Although the formation of a MBH from the gas concentration remains
purely speculative, the remainder of the overall picture rests on many
well-established aspects of galaxy dynamics.  It has long been known
\citep[e.g.][]{gerin_etal_88,sakamo_etal_99} that bars drive large
amounts of gas into galaxy centers, and \citet{fanali_etal_15}
emphasized that substantial inflow occurs during the process of bar
formation.  Work by others, \citep[e.g.][]{athana_92b, kim_etal_12a},
has indicated that the inflow often stalls at a gaseous ring having a
radius a few hundred parsec, which plausibly corresponds to observed
star-forming nuclear rings \citep{buta_95}.  Circum-nuclear
star-forming gaseous rings are preferentially found in barred
galaxies, leading to a correlation between bars and enhanced central
star formation \citep[e.g.][]{haward_etal_86, jogee_etal_05,
  mazzuc_etal_08, hao_etal_09, lin_etal_16}.

However, nuclear rings are not always formed in bar-driven flows, and can be
absent in cases where the central potential well of the galaxy is
relatively shallow, i.e.\ when the galaxy lacks a central massive
object.  We explain this dichotomy of behavior in section
\ref{sec:simulationresults}.

A strength of the model proposed by SM99 is that the process of
forming the MBH is self-limiting.  The build up of the central massive
object eventually causes gas inflow down the bar to stall at a nuclear
ring.  While some have argued \citep[e.g.][]{wada_04} for mild,
spiral-driven inflow within the ring, it must be at a much slower
rate, else the ring would not feature a gas density excess.
Furthermore, accretion onto the MBH would require the gas from the
ring to be carried inwards to $\la 1\%$ of the ring radius.  Thus the
fuel supply to the AGN is choked off by the formation of the nuclear
ring and MBH activity will decline.

SM99 included a growing massive central object in their models and
found, as is widely known \citep{norman_etal_96, she_sel_04},
that the bar may be weakened or even be dissolved as its mass
was increased.  \citet{kor_ken_04} argue that these secular internal
processes of bar-driven inflow causing the eventual dissolution of the
bar lead to the formation of ``pseudobulges'', which are thickened
central components with disk-like density profiles and velocity
distributions that may be formed without early galaxy mergers.

SM99, and also \citet{bou_com_02}, showed that on-going
accretion onto the disk of a galaxy may allow a second bar to form
later, but the massive compact object created in the first bar-forming
episode will cause gas inflow to stall at a nuclear ring, preventing
the MBH from becoming active again.

The purpose of the present paper is to determine the critical mass of the
central compact object needed to cause the bar-driven inflow to stall
at a nuclear ring, an issue that SM99 could not address since their
simulations lacked a gas component.  We use 2D hydro simulations in
realistic barred potentials to determine the central mass needed to
arrest the inflow at a nuclear ring.  We also estimate the likely
masses of the MBHs that could be formed before the nuclear ring is
established.

\begin{figure}[!t]
\epsscale{1.2} \plotone{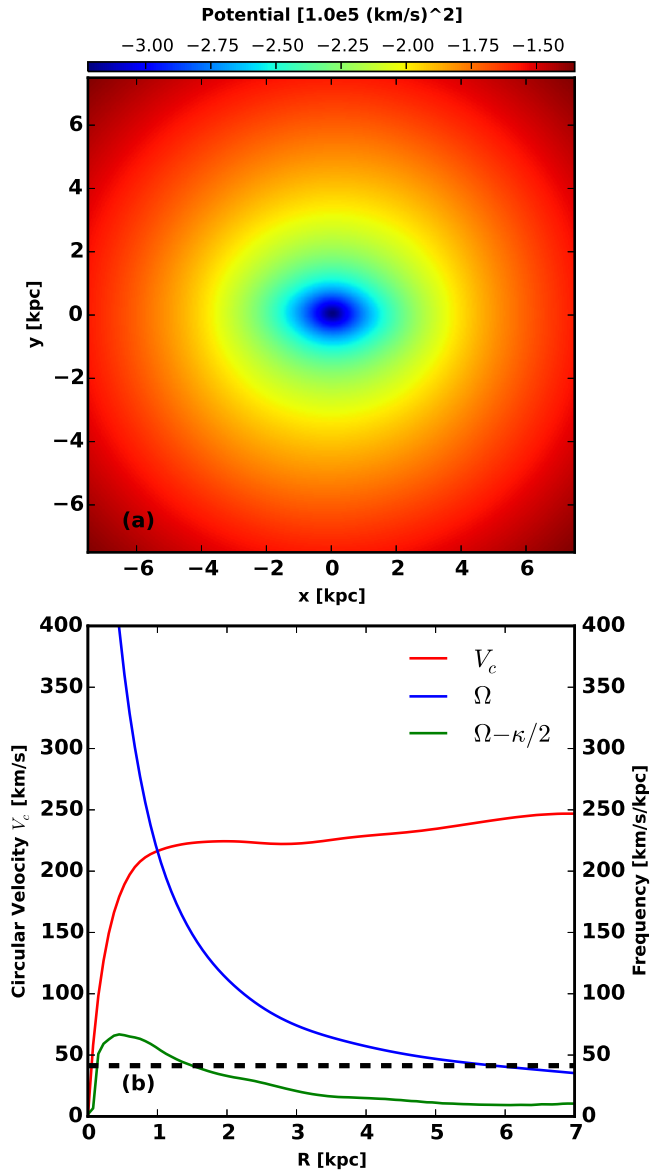}
\caption{Upper panel: contours of the potential of the stacked
  $N$-body model in a stationary frame.  The bar has a semi-major axis
  of $\sim3.5\kpc$ and is aligned with the $x$-axis.  Bottom panel:
  the rotation curve of the model together with curves showing the
  principal angular frequencies, that were determined from the
  azimuthally averaged central attraction.  The red line denotes the
  circular speed $V_c$, the blue line denotes $\Omega =V_c/R$, and the
  green line denotes $\Omega - \kappa/2$, where $\kappa^2 \equiv
  R^{-3}d(R^4\Omega^2)/dR$ \citep{bin_tre_08}.  The horizontal black
  dashed line shows the bar pattern speed, $\Omega_p =41.35\freq$ and
  the co-rotation resonance is at $R\sim5.7\kpc$.
\label{fig:rotcurve}}
\vspace{0.2cm}
\end{figure}

\section{Model Setup}

\subsection{Hydrodynamical Simulation}
\label{sec:simulationdetial}
We simulate gas flow in a rigid, rotating, non-axisymmetric barred
galaxy potential, focusing on the effects of massive central objects
on bar-induced inflow.  As the setup and the numerical methods used in
our simulations are very similar to those in \citet{li_etal_15}, we
give only a brief summary here.  We solve Euler's equations of ideal
hydrodynamics in the bar corotating frame using the grid-based MHD
code \textit{Athena} \citep{gar_sto_05, stone_etal_08, sto_gar_09}.
The gaseous disk is stirred by an external barred potential, described
in Section \ref{sec:generalpotential}, that is assumed to rotate
rigidly about the galactic center with a fixed pattern speed
$\mathbf{\Omb}=\Omb\mathbf{\hat z}$.  We adopt a 2D isothermal,
rotating gaseous disk with an initially uniform surface density of
${\Sigma}_0 = 15\Surf$, and neglect magnetic fields, and other
additional physics, except that we include self-gravity in one case.

We employ a uniform Cartesian grid with $4096\times4096$ cells
covering a box of size of $L=12.6\kpc$ in each direction.  Thus the
grid spacing is $\Delta x=\Delta y=3.1\pc$.  We adopt the
\textit{exact} nonlinear Riemann solver and outflow boundary
conditions at the domain boundaries for our hydrodynamic models.  The
importance of high spatial resolution and the exact Riemann solver has
already been demonstrated in previous work
\citep[e.g.][]{sorman_etal_15a, li_etal_15, few_etal_16}.  We choose
an effective isothermal sound speed of $\cs = 10\kms$ to describe the
mean velocity dispersion in molecular clouds, similar to earlier
studies \citep[e.g.][]{fux_99b, rod_com_08, kim_etal_12b}.  All the
models are run for a period of $1\Gyr$.

In one case only, we use Fourier transforms with periodic boundary
conditions to compute the gravitational potential of the gas as it
evolves.  This self-gravity term is added to that of the externally
applied bar potential.

\subsection{Gravitational Potential}
\label{sec:generalpotential}
Previous simulations of gas flow in barred potentials have generally
modeled the bar in one of two ways: as a rigid prolate spheroid
\citep[e.g.][]{athana_92b, kim_etal_12a} or by using the potential
from an $N$-body simulation in which a bar has formed through a disk
instability \citep[e.g.][]{fux_99a, shen_etal_10}.  It is easy to vary
the parameters of a prolate spheroid, especially one with an analytic
potential, but an $N$-body bar is both dynamically self-consistent and
is generally a better match to observed bars, as we show in Section
\ref{sec:robustness}.  Here we prefer to use the potential from an
$N$-body simulation.

\subsubsection{N-body Potential}
\label{sec:nbody}
We use the mid-plane potential from an $N$-body model created using
the \textit{GALAXY} code \citep{sellwo_14b}.  It is the model shown in
Figures 6.18 and 6.35 of \citet{bin_tre_08}.  The model began with an
exponential disk that was thickened with a ${\rm sech}^2(z/2z_0)$
vertical density profile.  The disk was embedded in a live halo that
was compressed \citep{sel_mcg_05} by the addition of the disk
from the initial density distribution with an isotropic distribution
function given by \citet{hernqu_90}.  The uncompressed halo had a
nominal mass 80 times and a scale radius 30 times those of the disk,
but any halo particles that would pass beyond a radius of $60R_d$
were discarded, reducing the halo mass to just over 20 times the disk
mass.  The disk was represented by 1 million particles and the halo by
2.5 million.  The mass of the disk was $3.0\times10^{10}\Msun$ and its
length scale was $1.5\kpc$.

We extracted the mass distribution of the simulated barred model at 26
moments over the period 225 to 250 dynamical times, where the
dynamical time is $5\Myr$.  As illustrated in Fig 6.35 of
\citet{bin_tre_08}, this period is well after the bar had formed,
buckled, and settled.  We stacked the separate mass distributions
after rotating each to a common bar major axis, assuming a steady
rotation rate at the best-fit pattern speed over this interval, and
derived the in-plane forces and potential from this time-averaged mass
distribution.

The top panel of Figure \ref{fig:rotcurve} shows the potential, in an
inertial frame, in the mid-plane of this mass model.  The bar has a
mean pattern speed of $\Omega_b=41.35\freq$ and bar strength parameter
$\Qb=0.40$ that implies a strong bar.\footnote{$\Qb$ is defined as the
  maximum ratio of the tangential force (mainly due to the
  non-axisymmetric bar potential) to the azimuthally-averaged radial
  force in the potential \citep[e.g.][]{com_san_81, comero_etal_10}.}
The lower panel of this Figure shows the corresponding rotation curve
and usual angular frequencies that were derived from the azimuthally
averaged central attraction.  The corotation radius is therefore at
$R_c \simeq 5.7\kpc$ and ${\cal R} \equiv R_c/a_B \simeq 1.6$, where
the semi-major axis of the bar, $a_B \simeq 3.5\kpc$, was estimated by
the method described in \citet{deb_sel_00}.

As usual, the hydrodynamical simulations start from a circular flow
pattern in the azimuthally averaged potential, and we compute the gas
flow in a frame that corotates with the bar.  To avoid subjecting the
gas flow to a sudden change, we gradually diminish the axisymmetric
potential to zero and substitute an increasing fraction of the
bisymmetric one over the first $100\Myr$.

\subsubsection{Massive central objects}
\label{sec:mco}
We find that a large amount of gas is quickly driven into a small
volume at the center of the model.  Since our simulations neglect many
physical processes, they cannot predict the fate of this gas.
However, SM99 argued that the principal dynamical consequence of the
accumulated gas would be to create a massive central object, which may
be composed of stars, gas, and a central MBH.  The precise nature of
the object is unimportant for the dynamics of the gas outside its
small radial extent.  In our models without gas self-gravity, we
therefore model the consequence of the inflow by adding a central
mass.  We also present one case where self-gravity is included to show
that the effect of mass accumulation in the center is well-captured by
the addition of a massive compact object.

We have employed three different density profiles (Plummer sphere,
Hernquist, and modified Hubble) to model the massive central object:
\begin{eqnarray}\label{eq.mcos}
{\rho}(r) &=& \left(\frac{3M_{\rm Plum}}{4{\pi}a^3}\right)
     \left(1+\frac{r^2}{a^2}\right)^{-5/2}, \nonumber \\
{\rho}(r) &=& \left(\frac{M_{\rm Hern}a}{2{\pi}r}\right)\frac{1}{(r+a)^3}, \\
{\rho}(r) &=& \rho_{\rm Hubb}\left(1+\frac{r^2}{a^2}\right)^{-3/2}.  \nonumber
\end{eqnarray}
Here $M_{\rm Plum}$ and $M_{\rm Hern}$ denote the total mass of the
central object by the corresponding density profile, while $\rho_{\rm
  Hubb}$ represents the object central density for modified Hubble
profile, and $a$ is the scale length of the central objects.  We
compute the gas flow patterns when this component is added to the
gravitational field of the $N$-body model, and compare the results
with cases where it is omitted.

\begin{figure*}[!t]
\epsscale{1.1} \plotone{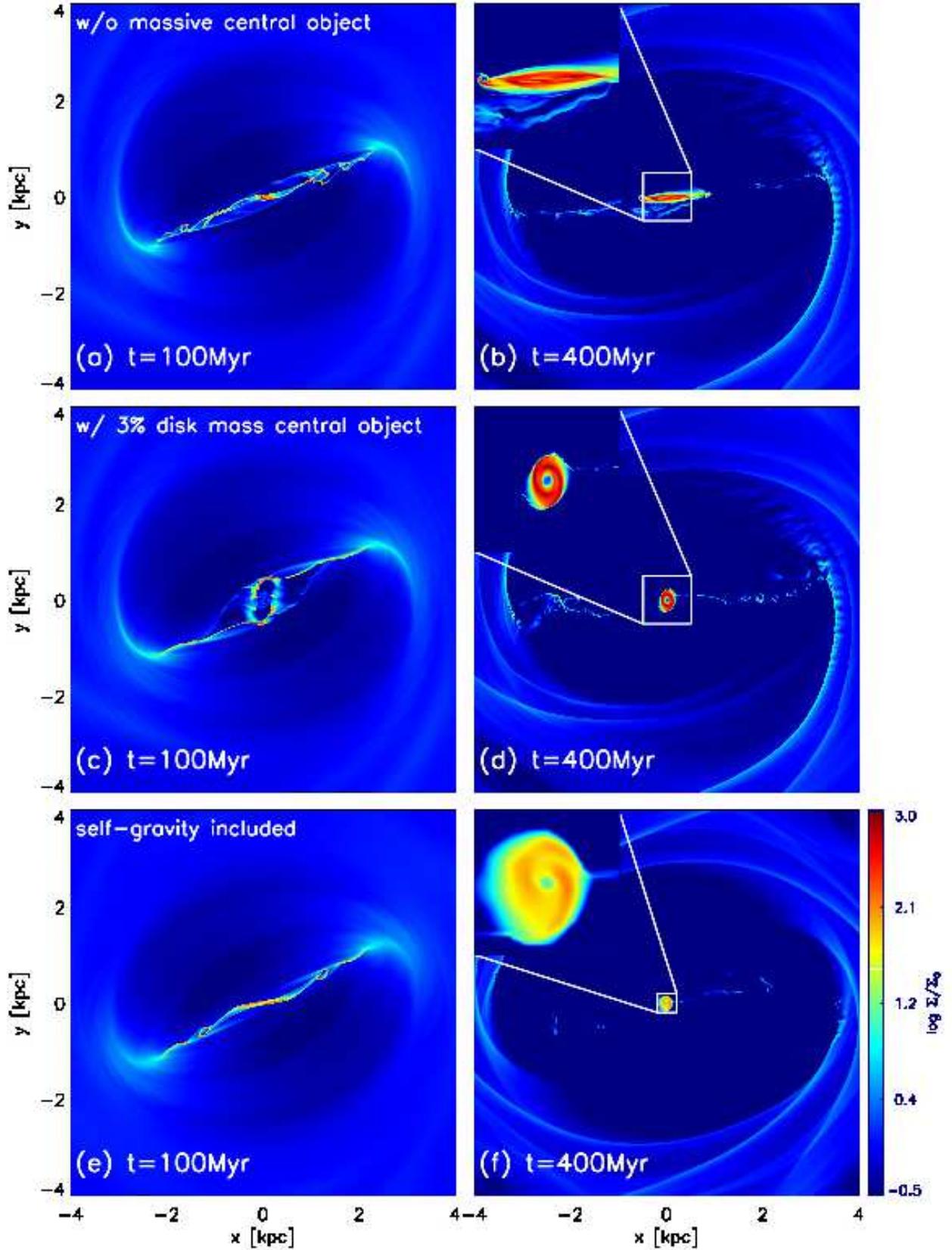}
\caption{The evolution of the gas surface density in models both
  without and with a massive central object.  The color scale is
  logarithmic.  Panels (a) and (b) are for the $N$-body model
  potential at two times, (c) and (d) illustrate the flow when a
  massive central object is included, while (e) and (f) show the bar
  model with gas self-gravity and no central object.  Note that the
  massive central object greatly alters the gas behavior in the
  central part ($R\le\sim1\kpc$), while the outer flow patterns
  ($R\ge\sim4\kpc$) are almost identical.
\label{fig:densecenter}}
\vspace{0.2cm}
\end{figure*}

\section{Two different gas flow patterns}
\label{sec:simulationresults}

Here we report simulations of the gas evolution in the potentials
described in Section \ref{sec:generalpotential}.  Since gas flows at
highly supersonic speeds relative to the potential everywhere except
close to the corotation resonance, strong shocks develop that are
indicated by the high density gas ridges (Figure
\ref{fig:densecenter}a).  Those within the bar form on the leading
side of the bar, as are typically found in all other work, and gas
flows rapidly towards the center.  This happens because gas loses both
energy, due to the shocks, and (on average) angular momentum, since it
is asymmetrically distributed with respect to the major-axis of the
potential.

\subsection{No central mass concentration}
The nature of the inner gas flow is strongly dependent on the
existence of an inner Lindblad resonance (ILR).  In the absence of
pressure forces, gas flows will settle onto stream lines that follow
nearly circular periodic orbits.  For a weak perturbing potential,
linear theory \citep{san_hun_76, bin_tre_08} predicts that the
orientation of a near circular closed loop orbit switches from
parallel to the bar on the side nearer to corotation to perpendicular
to the bar inside the ILR.  

Note that the condition $\Omb = \Omega - \kappa/2$ determines the
existence and locations of Lindblad resonances only for a bar of
infinitesimal amplitude.  Not only do the circular and epicycle
frequencies require generalization to action-angle variables in a bar
of finite amplitude, but the existence of resonances can be determined
only from orbit integrations -- see \citet{con_gro_89},
\citet{sel_wil_93} or \citet{bin_tre_08} for reviews.  Many families
of orbits have been identified in steadily rotating bar-like
potentials, even when motion is confined to a plane, but only two are
of importance here.  The main family of bar-supporting orbits, known
as $x_1$, is aligned parallel to the bar when viewed from a frame that
rotates with the bar.  However, a second family, known as $x_2$, that
is aligned perpendicularly to the bar often exists deep inside the
bar.  The change of orientation suggests that Lindblad's concept of a
resonance can be generalized to non-linear perturbations
\citep{van_san_82, li_etal_15}, and we hereafter extend the acronym
ILR to indicate the existence of the $x_2$ orbit family in
perturbations of finite amplitude.

Since the flow must have a unique velocity everywhere, pressure and
possibly shocks change the flow pattern where orbits intersect -- for
example in the region where orbit orientations switch from parallel to
perpendicular.  If this were to happen, we usually find that the
inflow stalls, and a dense, moderately eccentric, ring of gas builds
up where the $x_2$ orbits are found.

A necessary, but not sufficient, condition for the presence of the
$x_2$ family is that a weak bar of a given pattern speed should
possess two ILRs in the azimuthally averaged potential, as in the case
in our model, as shown in Figure \ref{fig:rotcurve}(b).  However, it
has long been known \citep{con_pap_80} that the possible
energy\footnote{More correctly, Jacobi constant, $E_J = E - \Omb L_z$
  which is an energy-like conserved quantity in a rotating
  non-axisymmetric potential.}  range of $x_2$ orbits narrows, and may
vanish entirely, as the strength of the bar perturbation is increased.

Indeed, no $x_2$ orbits exist in the strongly barred potential of our
original $N$-body model and therefore no ILRs were present in our
first model lacking a central mass concentration.  In this case,
therefore, gas was driven as far inward as we could resolve, as was
argued by SM99.  In Figure \ref{fig:densecenter}(a) the high density
gas ridges extend from $R\sim2.5\kpc$ to very close to the center,
indicating that gas is driven into the nuclear region.  Although, our
idealized hydrodynamic simulations cannot predict the ultimate fate of
the inflowing gas, it seems plausible that a high concentration of
stars will be formed, while some of the gas may connect to an
accretion disk to create a MBH.

The timescale for the inflow is also very short: during the first
$100\Myr$ when the bar is being established, a large fraction of the
gas has already been driven to the center.  By $t=300\Myr$, nearly all
the gas inside bar corotation radius has been driven to a highly
eccentric structure with corresponding streaming velocities.  The flow
pattern, which is shown at $t=400\Myr$ in Figure
\ref{fig:densecenter}(b), then becomes quasi-steady until the end of
the simulation mainly because we do not model the additional physical
processes that would probably consume the high density gas around the
center.

\subsection{Including a central mass concentration}
We wish to understand how the flow pattern is affected by the
accumulation of mass in the center.  In order to achieve this without
including gas self-gravity, we ran a series of additional simulations
in which we simply included the potential of central objects of
differing masses, described by one of the equations (\ref{eq.mcos}),
into that of the original $N$-body model.  The massive central object
was introduced at the beginning of each simulation and was not changed
while the simulation runs.

Periodic orbit studies in these modified potentials confirm the
existence of $x_2$ orbit family (i.e.\ an ILR) when we include a
central object of at least $\sim 3 \times 10^8\Msun$, which is
$\sim1\%$ of that of the galactic disk.  Although the bar is the same,
an ILR is enabled by the higher frequencies of the inner orbits in the
deeper potential well created by the central mass.

Figure \ref{fig:densecenter}(c) and (d) shows the gas flows at
$t=100\Myr$ and $t=400\Myr$ respectively when the central object was a
Plummer sphere with a scale length $a=100\pc$ and mass $M_{\rm
  Plum}=9.0\times10^8\Msun$, which is $3\%$ of the disk mass.  In
contrast to the run where no ILR is present (panels a and b), the high
density gas ridges curve around the center at their inner ends and do
not approach closer than $\sim 400\pc$ to the center, which is the
approximate radius of the ILR, see \S\ref{sec:cmcvary}.  The later gas
flow forms a ring-like structure with very little further inflow,
which is similar to nuclear rings often observed in barred galaxies.
Once an ILR is present, little, if any, of the bar-induced inflow can
reach the nuclear region, and the activity of the MBH must decline as
the immediately surrounding fuel is no longer being replenished by bar
inflow.

\subsection{Inclusion of gas self-gravity}
In order to test whether an artificially added massive compact object
adequately mimics the self-regulated process of mass inflow, we here
present a simulation that includes gas self-gravity and no externally
imposed massive central object.  

However, the inclusion of self-gravity is not straightforward because
an isothermal equation of state prevents the internal pressure from
rising as gas contracts under its own self-gravity, causing a runaway
density increase, as is well known.  Thus, naively including the
self-gravity term caused the central gas disk in our model to contract
continously with no more than a hint of a transient nuclear ring.
Changing to an adiabatic equation of state was not a solution, since
shocked gas remained hot, and the entire flow quickly ceased to be
supersonic, inhibiting almost all inflow of gas to the bar center.
The isothermal assumption is physically more reasonable, as gas in
galaxies dissipates energy efficiently, but dense gas also fragments
to form stars, with energy feedback on small-scales that is
challenging to model.  The galaxy formation community adopts rules to
try to capture the unresolvable ``sub-grid physics'' of star formation
\citep[e.g.][]{Scann_12}.  These rules are themselves no more than
physically-motivated guesses and, for our purpose, are both too time
consuming and complicated, since we do not need to track the highly
local re-expansion and pollution of the gas, the ages and chemistry of
the stars, etc.

We therefore adopted a simpler approach.
\begin{itemize}
\item We model a steady rate of star formation by continuously
  removing mass from the gas and replacing it with a growing rigid
  mass distribution to represent the gravitional attraction of the
  newly-formed stars.  Starting at $t=150~$Myr, when some gas had
  already accumulated in the center, we removed a fraction of gas
  having a 2D exponential density profile from the gaseous disk at
  every time step, while adding the gravitational attraction of the
  removed gas to the rigid background potential.  We chose a
  scalelength of 10~pc for the exponential profile, and froze 1\% of
  the surface density at center at each time step.  Our results were
  insensitive to reasonable variations of these parameters, but a much
  lower freezing fraction caused too slow a rise in the mass in
  stars.

\item We found that the nuclear gas disk was destabilized by a minor
  asymmetry in our $N$-body model, which created a rotating, mildly
  lop-sided component to the rigid forcing potential.  We therefore
  imposed 4-fold symmetry by reflecting the potential about the $x$-
  and $y$-axes and averaging.

\item We increased the initial gas surface density to
  $30~$M$_\odot$~pc$^{-2}$, resulting in a gas fraction that is $\sim
  10\%$ of the stellar disk mass inside the bar co-rotation radius.
  Though large for present-day galaxy disks, such a gas fraction is
  probably on the low side for galaxies in the early universe.

\end{itemize}

\begin{figure}[!t]
\epsscale{1.1} \plotone{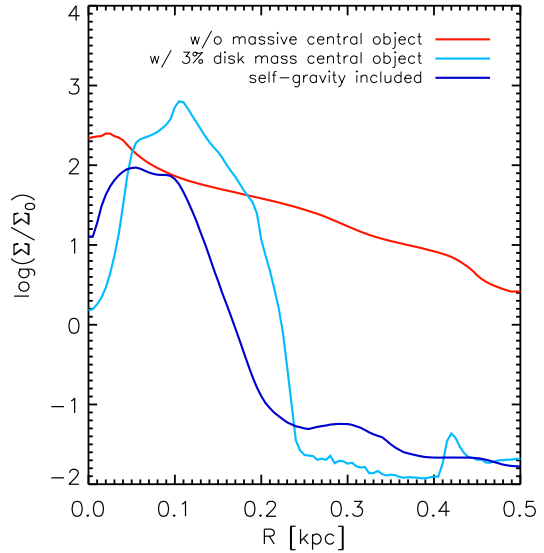}
\caption{The gas density profile along the bar minor axis at
  $t=400\Myr$ in the three simulations shown in Figure
  \ref{fig:densecenter}. The red curve is for the run with no central
  object, the cyan curve is for the run with a rigid central object,
  and the blue curve is for the run with self-gravity.}
\label{fig:denseprofile}
\vspace{0.2cm}
\end{figure}

\begin{figure*}[!t]
\epsscale{1.2} \plotone{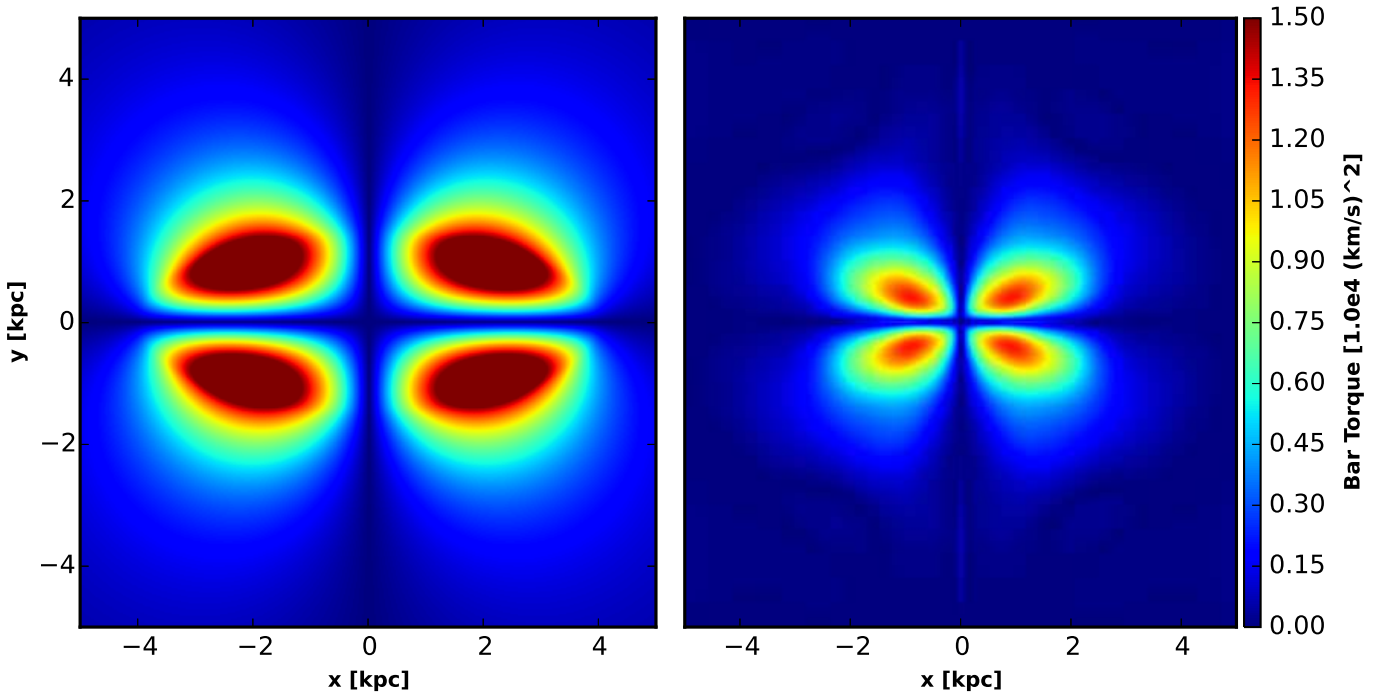}
\caption{The bisymmetric torque $|\mathbf{R} \times \mathbf{F}|$ for an
  $n=1$ Ferrers prolate spheroid (left panel) and for the $N$-body bar we
  used (right panel).  The spheroid has semi-major and semi-minor axes
  of $4.0\kpc$ and $1.5\kpc$, similar to the shape of the $N$-body
  bar, and the bar strength parameter $\Qb=0.40$ for both bars.
\label{fig:diffbars}}
\vspace{0.2cm}
\end{figure*}

The flow pattern in this case is shown in panels (e) and (f) of
Figure~\ref{fig:densecenter}. The evolution for the first 100 Myr
resembles that shown in (a), but panel (f) shows a ring-like structure
by $t= 400~$Myr in a manner that more closely resembles that shown in
panel (d), and the ring persists until the end of the simulation
(1~Gyr).  In this case, the mass of frozen gas is $7.1 \times
10^8~$M$_\odot$ at $t=400~$Myr, rising slightly to $7.8 \times
10^8~$M$_\odot$ by $t=1~$Gyr, which is approximately 2.6\% of the disk
mass.  Note that the rate of star formation in this bar-driven nuclear
starburst is less than 3~M$_\odot$~yr$^{-1}$, when averaged over the
interval 150 -- 400~Myr, in line with present-day estimates for barred
galaxies \citep[e.g.][]{Davi17}.  Thus the mass of gas that reaches
the central parts in this model is comparable to that we assumed for
the rigid mass concentration in the middle row of
Figure~\ref{fig:densecenter}, and is sufficient to cause an ILR that
stalls the inflow, as discussed above.

Figure~\ref{fig:denseprofile} shows the azimuthally averaged gas
density profile in all three simulations shown in
Figure~\ref{fig:densecenter}.  The red curve is for the run with no
central object for which the density rises all the way to the center.
The cyan curve shows the ring-like feature that formed in the run with
a rigid central object.  The blue curve is for the run with
self-gravity, which also presents a ring-like structure.  These
results clearly show that, when self-gravity is included, the
bar-driven flow creates a central mass concentration that, once
established, causes the subsequent inflow to stall at $\sim 100~$pc
from the center.

\section{Tests}
\label{sec:robustness}
The large-scale gas flows presented in the previous section support
the evolution scenario proposed in SM99.  They suggest that bars may
contribute to AGN activity for only a very short time due to the rapid
response of gas to the barred potential.  Most inflow may occur during
the bar formation episode, as reported by \citet{fanali_etal_15}, and
nuclear activity should cease soon after the bar and the ILR are fully
established.

In this section, we present tests to demonstrate that our results are
affected neither by the barred model nor by the density profile of the
massive central object we adopt.  However, we do find some evidence
for mild numerical diffusion at $R \la 100\pc$.

\subsection{Bar model}
The gas flow patterns are sensitive to the adopted potential of the
bar because its shape and mass determine the torque, $|\mathbf{R}
\times \mathbf{F}|$, acting on the gas; here $\mathbf{R}$ and
$\mathbf{F}$ are the position and gravitational force vectors,
respectively.  Figure \ref{fig:diffbars} compares the distributions of
the bar torque for our $N$-body bar with that of a similar size $n=1$
Ferrers prolate spheroid that has a similar axis ratio $b:a = 3:8$,
and almost the same $\Qb$ parameter.  We see that the forces from the
$N$-body bar differ significantly from those of the spheroid model in
the sense that the high torque regions in the $N$-body bar form more
of an X-shape and are more centrally concentrated.  Not only is this
very similar to the corresponding maps derived from the photometry of
barred galaxies \citep{but_blo_01}, but the stronger torques at small
radii can drive gas closer to the center.

We find that the gas flow patterns in the $N$-body bar differ slightly
from those in the Ferrers potential, which we adopted in our previous
work \citep{li_etal_15}.  A gaseous ``inner ring'' that lies on the
edge of the bar is commonly seen in the simulations using a prolate
spheroid \citep[e.g.][]{athana_92b, kim_etal_12b}, while such a
feature has not been found in observations\footnote{It is worth noting
  that the power-law density prolate spheroid bar model used in
  \citet{binney_etal_91} is more similar to an $N$-body bar both in
  the torque distribution and in the resulting gas flow pattern.}.  In
addition, \citet{fragko_etal_16} argued that boxy/peanut (B/P) bulges
reduce the amount of gas reaching the central regions.  The formation
of B/P bulges is a natural consequence of an $N$-body bar, while a
prolate spheroid is not a close match to a B/P shape.  We therefore
conclude that gas flow patterns in an $N$-body bar may better match
those in galaxies.

Since our bar model is somewhat ``slow'' (${\cal R} \sim 1.6$) we also
tested the ``fast'' bar from \citet{shen_etal_10}, for which ${\cal R}
\simeq 1.2$.  We found that the gas flow patterns are almost identical
to those presented in \S\ref{sec:simulationresults}, suggesting that
the critical central mass of 1\% disk mass is a robust result for
different $N$-body bars.

\begin{figure}[!t]
\epsscale{1.2} \plotone{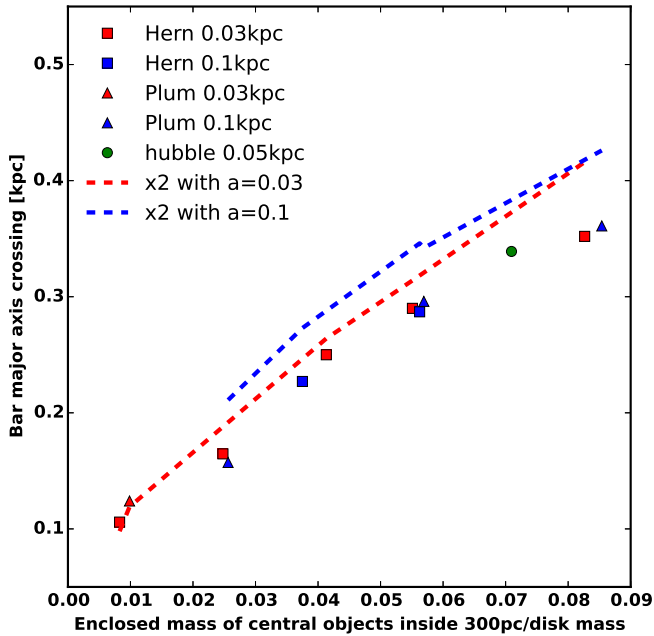}
\caption{The ring outer radius at $t=400\Myr$ as a function of the
  mass of the central object enclosed mass inside $300\pc$, expressed
  as a fraction of the disk mass.  The squares are from the
  simulations using Hernquist profile and the triangles are using
  Plummer sphere.  Red and blue represent the models having a scale
  length of $0.03\kpc$ and $0.1\kpc$.  The green circle is from a
  simulation using modified Hubble profile with a scale length of
  $0.05\kpc$ and an enclosed mass of $2.0\times10^9\Msun$ at
  $R=300\pc$.  The dashed lines show the major axis crossing of the
  largest non-intersecting $x_2$ orbits for two different values of
  the bulge scale parameter $a$.}
\label{fig:diffcenters}
\vspace{0.2cm}
\end{figure}

\subsection{Different central masses}
\label{sec:cmcvary}
We found (\S\ref{sec:simulationresults}) that $x_2$ orbits (or an ILR)
that cause the gas inflow to stall in a nuclear ring are present only
in models that include a massive central object.  Here we study the
extent to which the choice of parameters of the massive central object
affects the inner part of the flow pattern.  We compare results from
two density models: a Hernquist profile with a scale length of $30\pc$
to represent a cusped central object, and a Plummer sphere with a
scale length of $100\pc$ to represent an object with an harmonic core,
and vary the mass for each over the range 1\% to 10\% of the disk
mass.  In each case we estimated both the outer radius of, and
mass of gas in, the nuclear ring.

In the absence of thermal pressure or other forces acting on the gas,
the stream lines will follow periodic orbits that do not intersect.
However, shocks will remove gas from orbits that do intersect.  The
highest energy $x_2$ orbits are skinnier than those of lower energy
and intersect with some of them.  Thus, the gas must settle onto the
lower energy $x_2$ orbits that are smaller and rounder than those of
the highest energy.  We therefore expect the nuclear ring to form
approximately near the inner edge of the range of $x_2$ orbits,
although the correspondence may not be exact since pressure forces may
also affect the motion of the gas to a small extent.

The variation of the ring radius as a function of the mass of the
central object enclosed within $300\pc$ is plotted in
Figure~\ref{fig:diffcenters}.  We find that a more massive central
object with a smaller scale length results in a larger ring outer
radius.  We conclude that the radius where the flow stalls depends
most strongly on the mass of the central object and is less sensitive
either to the density profile or to its scale length over the tested
range.  The dashed lines plot the central distance at which the
largest $x_2$ orbit that does not intersect other $x_2$ orbits crosses
the bar major axis for two different values of the bulge scale
parameter.  It is clear from this Figure that rings form where gas can
settle onto non-intersecting $x_2$ orbits.

We also find that the total mass of the ring is about the same ($\sim
4.5\times10^8\Msun\sim1.5\%$ disk mass) in all the models, regardless
of the masses or scale sizes of the central object.  Probably this is
simply a consequence of holding the bar size, pattern speed, etc., as
well as the gas density, fixed in all our simulations.  See
\citet{li_etal_15} for more discussion.

Note that the lowest central density Plummer model, with 1\% of the
disk mass and $100\pc$ scale length, did not form a nuclear ring
(i.e.\ gas flowed to the very center), but a ring did form at the same
mass when we reduced its length scale to $30\pc$.  Considering the
intrinsic scatter in observed bar properties and the fact that nuclear
star clusters are very dense, we conclude a massive central object
with about $\sim1$\% disk mass is enough to prevent further inflow to
the MBH.

\begin{figure}[!t]
\epsscale{1.2} \plotone{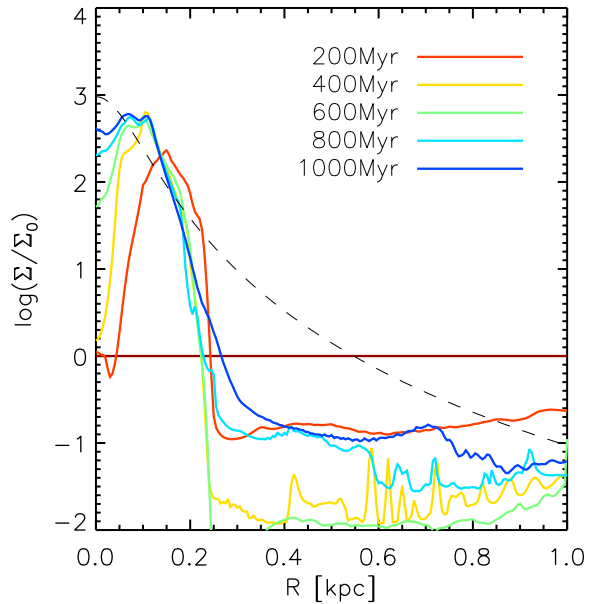}
\caption{Normalized azimuthally averaged gas surface density inside
  $1\kpc$ in the simulation using a Plummer sphere with a mass of
  $9.0\times10^8\Msun$ and a scale length of $0.1\kpc$.  The colors
  represent gas density at different time.  The initial density
  distribution is shown by the horizontal deep red line.  The ring
  becomes a disk at around $600\Myr$.  We also draw the surface 
  density of a Plummer sphere with a scale radius of $0.1\kpc$ and a 
  mass of $4.5\times10^8\Msun$ using the black dashed line.
\label{fig:innerboundary}}
\vspace{0.2cm}
\end{figure}

\subsection{The inner boundary of the ring}
The gaseous nuclear ring formed due to the massive central object
usually presents an inner boundary, and we find little gas inside.
However, we observe that the inner boundary shrinks with time and
eventually the ring may fill to become a disk (Figure
\ref{fig:innerboundary}).  Since we wish to be clear whether the
processes we include in our simulations do or do not allow gas to
reach the center and accrete onto a MBH, we have tested whether the
filling of the ring is physical or numerical.

We performed simple simulations that began with a rotating isothermal
gaseous ring, having a sharp inner boundary, in the axisymmetric
potential of an exponential disk.  We varied the spatial resolution
and the sound speed and studied whether the boundary blurred over
time.  We found that simulations with at least 30 grid points inside
the inner boundary maintained a sharp edge, while with fewer than 20
grid points inside the initial edge the ring gradually diffused
inwards, filling the hole completely after a few tens of orbit
periods.  Thermal effects on spreading the boundary were negligible as
long as the rotation velocity was large compared with the sound speed.

We have also used the static mesh refinement (SMR) technique to better
resolve the central regions of the simulations using the bar model,
finding that the ring lasts for a longer time before filling in the
SMR runs.  We therefore conclude that the spreading of the inner
boundary in our simulations is caused by numerical diffusion, and that
simple hydrodynamics alone would predict that an ILR would perfectly
cut off the supply of gas to a MBH in the galactic center.  Note it
is possible that other physical processes could intervene to bring gas
from the nulcear ring to the MBH as we discuss below.

\section{Discussion}
\label{sec:discussion}
\subsection{Mass of the central BH}
In order to form a MBH, the angular momentum of gas in the disk of a
galaxy must be reduced by many orders of magnitude.  Accretion disks
and dusty tori are invoked for the last stages
\citep[e.g.][]{krolik_99}, but gas must be brought to a radius of a
few parsec before they can take over.  We have found, in common with
most other work, that a large fraction of the gas in the bar region is
driven very close to the center in the first $\sim 100\Myr$.  We also
found from our run with self-gravity, Figure~\ref{fig:densecenter}(f),
that inflow is stalled at a well-established nuclear ring $\sim
400\Myr$ after the start.  Thus abundant gas is present in the nuclear
region for some $\sim 300\Myr$.

Here we make no attempt to calculate the evolution of this gas
concentration, and confine ourselves to speculation.  A MBH may form
through a cascade of instabilities to ever smaller scales, along the
lines of the models proposed by \citet{shlosm_etal_89} and
\citet{hop_qua_10}, but without the pre-existing MBH.  Alternatively,
vigorous star formation in such environment could lead to the runaway
growth of a massive star \citep{krumho_15}.

Whatever the initial mass of the BH, the rich reservoir of gas in the
nucleus will enable it to grow rapidly.  In order to make a very rough
estimate of the mass of the MBH that may be formed, we assume that the
proto-BH accretes $\sim 0.01\Msun$ of gas each year, say, which would
lead to a not unreasonable final mass of $\sim 3 \times 10^6\Msun$.
The remaining $\sim 99\%$ of the gas in the nuclear region would make
a nuclear star cluster or small pseudo-bulge, and some could be
ejected through stellar and/or AGN feedback.

\subsection{Bars and AGN}
Fuel must be supplied to the accretion disk surrounding a central MBH
in a galaxy in order for it to become active.  We have shown that
bar-driven gas inflow stalls at a nuclear ring when the galaxy hosts a
massive central object that exceeds 1\% of the disk mass.  In our
idealized simulations, the nuclear ring makes a ``watertight'' barrier
that prevents the stalled gas from reaching the central engine.  In
reality, stellar feedback, magnetic fields, self-gravity, etc.\ may
allow some slow leakage of gas from the ring to the nuclear region,
but since the nuclear ring radius is many hundreds of times the scale
of the accretion disk, only gas that has somehow shed over 99\% of its
angular momentum before reaching the accretion disk could fuel the
MBH.  These issues were reviewed by \citet{jogee_06}.

Thus, except in very earliest stages of galaxy assembly before the
massive central object has been created, our models predict little or
no connection between the fueling of AGN and the presence of a bar in
the disk; i.e.\ whatever causes AGN activity in galaxies should be
largely unaffected by whether the galaxy hosts a bar.  Furthermore,
any possible relation between a bar and a MBH is likely to be erased
by the possible self-destruction of the first bar and the formation
of a new bar.

This prediction seems consistent with the findings of many
observational studies that have examined the possible connection
between large-scale bars and feeding of MBHs.  Some authors have
concluded that AGN activity is mildly enhanced in barred galaxies
\citep[e.g.][]{hao_etal_09, oh_etal_12, alonso_etal_13,
  gallow_etal_15} but others have not \citep[e.g.][]{kor_ho_13,
  cheung_etal_15}.  An additional study by \citet{cister_etal_15}
used {\it Chandra} X-ray data to identify AGN and HST imaging to
examine the morphology of galaxies out to $z\sim0.84$ and also
concluded that the presence of a bar had no influence on the strength
of AGN activity.  This body of work therefore suggests that, while a
tendency for bars to cause a mild increase in AGN activity is not
fully excluded, it is clearly not a strong effect.

Many of these papers examined low redshift ($z \la 0.05$) samples of
galaxies.  Even the studies by \citet{cheung_etal_15} and
\citet{cister_etal_15} out to $z \ga 1$, used observations at an epoch
that is long after our predicted early connection between bars and the
creation of a MBH has been erased.  An observational study at much
higher redshift to test for a possible connection between the
morphology of forming disks \citep[see][for a discussion of the
  detectability of bars]{Erwin_17} and AGN activity seems well beyond
what is technically feasible today, and may even be beyond the reach
of the James Webb Space Telescope.

In other theoretical work, \citet{shlosm_etal_89} proposed that
nuclear gaseous rings could become dynamically decoupled from the bar
that formed them, enabling a cascade of bar instability events that
could drive gas closer and closer to the MBH, a picture that was
refined by \citet{hop_qua_10} using multiscale SPH simulations.  It
seems hard to reconcile this predicted behavior with the weak, or
non-existent, correlation between bars and AGN.

\subsection{Bar formation}
We assume that a bar forms quickly as the rotationally supported disk
is being assembled, which is likely since disks are chronically
unstable.  Modern simulations \citep{athana_02, sah_nab_13,
  polyac_etal_16} have shown that bars form far more readily in live
halos than in rigid, and that bar formation seems inevitable in
rotationally supported disks of even quite low mass
\citep{ber_sel_16}.  These models also find that bars can form in
halos with central density cusps, as happened in the model we employ
here, and a harmonic core is therefore not required for bar formation.
The only requirement to seed a MBH in the SM99 model is that the
initial mass distribution of the disk plus halo should not include a
massive central object that would cause the gas flow to stall at a
radius of a few hundred parsec.

Observations of forming galaxies at $z \ga 2$ reveal that they
generally have a turbulent, irregular clumpy appearance
\citep{elmegr_etal_07}.  However, they do seem to have significant
rotation \citep{shapir_etal_08} and bar formation is hard to prevent
even if the underlying mass distribution were as clumpy as the light
\citep[e.g.][]{du_etal_15} which seems
unlikely as the bright spots are believed to be areas of intense star
formation.

\subsection{Relation to the merger model of BH growth}
SM99 proposed that a moderate mass MBH should be formed as the disk is
being assembled and forms its first bar.  The activity is
self-limiting because the build-up of a central concentration causes
the gas inflow to stall at a circum-nuclear ring, thereby starving the
central engine of further fuel.  These precursor MBHs are required to
power quasar activity during subsequent mergers.

Furthermore, as reviewed in the introduction, the MBHs formed by the
mechanism proposed by SM99 are required as modest central engines of
Seyfert activity in galaxies lacking a significant classical bulge
component.  The absence of a substantial classical bulge is generally
believed to indicate that the host galaxy has not experienced
significant mergers as it was being assembled, as was stressed by
\citet{kormen_etal_10}.

\subsection{Limitations}
Our simulations carefully compute the 2D flow of an ideal, isothermal
gas, in a fixed-potential.  But we neglect many complicating
physical processes that might affect the gas flow and the ultimate
fate of the gas accumulated in the galactic center.

We justify this approach because the sole science question we wished
to address was how the mass of the accumulated gas in the center
affects the subsequent flow pattern.  The inflow happens so
efficiently and quickly that star formation, feedback, and other
``gastrophysics'' can scarcely have time to affect the outcome, as
\citet{fanali_etal_15} argued.  The main effect is that the mass of
gas that accumulates in the center alters the subsequent flow; whether
the gas inside a few hundred parsec goes on to form stars and/or an
MBH will not change this result.

Other studies have shown that increasing the sound speed
\citep{kim_etal_12a}, including magnetic fields \citep{kim_sto_12},
gas self-gravity \citep{wada_04}, or star-formation and stellar
feedback \citep{izumi_etal_16} could enable a moderate inflow to MBHs,
even after the nuclear ring has developed.  But these processes will
influence the flow on a time scale that is long compared with that on
which the ring was created, and would have little affect on the
initial rapid growth of the MBH or the establishment of the nuclear
ring (or ILR).  On the other hand, any stellar and AGN feedback will
likely hasten the clearance of gas around the MBH, and may further
shorten the duration of AGN activity.  Therefore we consider that
inclusion of these sub-grid physical processes would not qualitatively
change our findings.

\section{Summary}
We have performed hydrodynamical simulations of gas flows in a
realistic $N$-body barred galaxy model that lacks a classical bulge.
We found two distinct flow patterns depending upon the presence of a
sufficiently massive central object.  Without a central object, gas
could be driven by the bar down to the very center, while the flow
stops at a few hundred parsecs away from the center when the central
object of at least $\sim1\%$ disk mass is included.  We have shown
that the ring forms near an inner resonance, where the higher orbital
frequencies cause the gas response to be perpendicular to the bar
major axis -- the generalization of the ILR from linear theory.  By
including self-gravity of the gas, and mimicking the formation of
stars in a simplified way, we also showed that a central object of
this mass could be created by the gas inflow itself.

We argue that moderate mass MBHs could be created in forming disk
galaxies due gas inflow driven by the first bar in the disk.  The
activity of these MBHs is self-limiting because the build-up of a
massive central object stalls the inflow at a nuclear ring, thereby
bringing about the end of rapid growth of the MBH.  We estimated the
mass of the MBH formed during such a process to be
$\sim10^6$--$\sim10^7\Msun$, which is typical of the masses of MBHs
required for Seyfert activity in nearby galaxies having pseudobulges,
but little or no classical bulge.  The bar might also be destroyed if
the central object is massive enough.  Gas flows in any bar formed
subsequently will stall at a nuclear ring, due to the continued
existence of the massive central object, preventing accretion onto the
MBH and leaving at most a weak correlation between bars and AGN today.

\acknowledgments We thank an anonymous referee for some very
constructive criticism that helped us to strengthen the paper, and
also Luis Ho for valuable comments.  The research presented here was
partially supported by the 973 Program of China under grant
no. 2014CB845700, by the National Natural Science Foundation of China
under grant nos.11333003, 11322326, and by a China-Chile joint grant
from CASSACA. A China Scholarship Council award (CSC, No.201604910905)
supported ZL during a six-month visit to Rutgers University.  We also
acknowledge support from NSF grant AST/12117937 to JAS and a {\it
  Newton Advanced Fellowship} awarded by the Royal Society and the
Newton Fund to JS, and from the CAS/SAFEA International Partnership
Program for Creative Research Teams.  This work made use of the
facilities of the Center for High Performance Computing at Shanghai
Astronomical Observatory.


\end{document}